\documentclass[aps,prd,twocolumn,showpacs,amsmath,amssymb]{revtex4}
\usepackage[utf8]{inputenc}
\usepackage[english]{babel}
\usepackage[dvips]{graphicx}

\newcommand{\om} \omega   \newcommand{\Om} \Omega
\newcommand{\eps} \epsilon
\newcommand{\be} {\begin{equation}}
\newcommand{\ee} {\end{equation}}
\newcommand{\ba} {\begin{eqnarray}}
\newcommand{\ea} {\end{eqnarray}}
\def\lrD{\mathrel{{\cal D}\kern-1.em\raise1.75ex\hbox{$\leftrightarrow$}}}
\def\lr #1{\mathrel{#1\kern-1.25em\raise1.75ex\hbox{$\leftrightarrow$}}}

\begin{document}

\title{Signatures of trans-Planckian dispersion in inflationary spectra}

\author{Jean Macher}
\email{jean.macher@th.u-psud.fr}
\affiliation{Laboratoire de Physique Th\'eorique, CNRS UMR 8627, B\^at. 210, Universit\'e Paris-Sud 11, 91405 Orsay Cedex, France}
\author{Renaud Parentani}
\email{renaud.parentani@th.u-psud.fr}
\affiliation{Laboratoire de Physique Th\'eorique, CNRS UMR 8627, B\^at. 210, Universit\'e Paris-Sud 11, 91405 Orsay Cedex, France}

\date{\today}

\begin{abstract}
The primordial spectra are calculated using dispersion relations which deviate from
the relativistic one above a certain energy scale $\Lambda$.
We determine the properties of the leading modifications 
with respect to the standard spectra when $\Lambda \gg H$, where $H$
is the Hubble scale during inflation.
To be generic, we parameterize the lowest order deviation 
from the relativistic law by $\alpha$, the power
of $P/\Lambda$ where $P$ is the proper momentum.
When working in the asymptotic vacuum,
the leading modification scales 
as $(H/\Lambda)^\alpha$ for all $\alpha$,
except for a discrete set where 
the power is higher. Moreover, this modification is robust against
introducing higher order terms in the dispersion relation. 
We then algebraically deduce the modifications of scalar and tensor
power spectra in slow roll inflation
from modifications calculated in de Sitter space.
The modifications do not exhibit oscillations
unless the dispersion relation induces some non-adiabaticity
near a given scale.
Finally, we explore the much less studied regime where $H$
and $\Lambda$ are comparable.
Our results indicate that the project of reconstructing the inflaton potential
cannot be pursued without making some hypothesis about the dispersion relation 
of the fluctuation modes.
\end{abstract}

\pacs{98.80.Cq, 98.70.Vc}

\maketitle

\section{Introduction}

In agreement with the observed temperature anisotropies in the 
Cosmic Microwave Background~\cite{Komatsu:2008hk}, 
inflation predicts an almost scale invariant 
spectrum of primordial density 
perturbations~\cite{Mukhanov:1981xt,Mukhanov:1992me}.
These adiabatic perturbations arise from the amplification 
of vacuum fluctuations of linearized metric and inflaton perturbations.
This mechanism relies on a semi-classical description, {\it i.e.},
quantum fields propagating in a background spacetime. However in most models, 
the inflationary phase lasts so many {\it e}-foldings
that the fluctuations we 
today observe stem from vacuum fluctuations characterized by wavelengths much shorter than the Planck length at the 
onset of inflation. In this regime the semiclassical description is no longer trustworthy.
It is therefore of importance 
to find out to what extent the inflationary
predictions depend on the physics that takes place above (or near) the Planck scale.

In the absence of a theory of quantum gravity, 
there is no obvious way to address this question.
In fact several approaches have been adopted. 
The first one is directly inspired by 
what was done in black hole physics 
where a similar problem occurs~\cite{Jacobson:1991bh,Unruh:1994je,Brout:1995wp}:
to test the robustness of the predictions,
one introduces some dispersion above a certain energy scale $\Lambda$,
and studies the sensitivity of the power spectrum to 
this modification~\cite{Martin:2000xs,Niemeyer:2000eh}.
It was shown that the properties of the power spectrum are robust, 
{\it i.e.}, the deviations with respect to the standard spectrum
are small whenever the adiabaticity of the evolution is 
preserved and $\Lambda$ taken well above the Hubble scale $H$ 
during inflation~\cite{Niemeyer:2001qe}.
However, the precise relationship between the
modifications of the dispersion relation and the induced modifications
of the spectrum was not obtained in the general case.
To get a {\it generic} estimate of the modifications
a simplified approach was proposed in Ref.~\cite{Danielsson:2002kx}. 
Instead of introducing dispersion around the scale 
$\Lambda$, the vacuum state was imposed when the momentum $P=\Lambda$
rather than in the asymptotic regime $P \to \infty$
 as done when
working in the Bunch-Davies vacuum. 
A third approach based on an effective action obtained by integrating out heavy degrees of freedom of characteristic mass $\Lambda$, was also used in  Ref.~\cite{Kaloper:2002uj} and a different estimate of the
modifications was obtained.

If all approaches agree on the fact that the inflationary 
predictions are robust when $\Lambda/H \gg 1$, there is indeed no
agreement about the general properties of the modifications
of the spectra.
In particular, there is no agreement concerning the power of $H/\Lambda$
which characterizes their amplitude. In Ref.~\cite{Danielsson:2002kx} it was 
argued that the deviations should {generically} 
be first order in $H/\Lambda$, whereas
in the effective lagrangian 
approach it was argued that the
corrections should be (at least) second order and contain
only even powers of $H/\Lambda$.
Moreover in a reformulation of the model of \cite{Danielsson:2002kx}
where the state imposed when $P = \Lambda$ is the (properly defined)
instantaneous adiabatic vacuum~\cite{Niemeyer:2002kh}, 
it was shown that the corrections are third order.
As of the first method based on dispersion,
as we said, no general result seems to have been obtained. 

In addition, besides the question of the amplitude of the deviations,
there is no agreement either on the generic character of the rapid oscillations which have been found 
in~\cite{Easther:2001fz,Niemeyer:2002kh,Martin:2003kp} 
(but not in~\cite{Kaloper:2002uj}), 
confronted with observational data in~\cite{Martin:2003sg,Martin:2004iv},
and criticized in \cite{Campo:2007ar}.

In the present work, we aim to settle the questions concerning the
amplitude of the deviations and the presence of fast oscillations when using dispersion,
and still working with the asymptotic (Bunch-Davies) vacuum.
To get results which are not bound to a particular dispersion
relation (or to a particular class thereof),
we parameterize the {\it first} deviation with respect 
to the relativistic law by a scale $\Lambda$ and a power $\alpha$
in the following way
\be
\Omega^2  = F^2(P^2) = P^2 \left( 1 \pm \left({P}/{\Lambda}\right)^\alpha 
+ O\left(\left({P}/{\Lambda}\right)^{ 2\alpha}\right) \right) \, .
\label{NL}
\ee
The sign determines whether the propagation is
super- ($+$) or subluminous ($-$). 
As in former works, 
the preferred frame which is used to define $\Omega$ and $P$
is taken to coincide with the cosmological frame: 
$\Omega$ is thus the proper frequency and $P$ the norm
of the spatial momentum as measured by comoving observers. 
In these models the isotropy and the homogeneity of FLRW are preserved,
but the {\it local}~\cite{Jacobson:1991bh,Jacobson:1996zs} Lorentz invariance 
has been broken. (This is not the case in~\cite{Kaloper:2002uj}.) 
It should also be noticed that these models respect a modified 
(weak)
Equivalence Principle~\cite{Parentani:2007dw}
in that, when considering high momenta (in the 
preferred frame) and neglecting the gradients of the metric,
the physics is the same as in Minkowski space (in the preferred rest frame).

To separate the modifications due to the dispersion above the scale $\Lambda$ 
from those governed by slow-roll parameters,
we first compute the deviations of the spectrum of a scalar field 
propagating in de Sitter space and then show how these determine, by simple substitutions,
those of scalar and tensor modes in slow-roll inflation.

In Section~\ref{modifspec}, to start the analysis, we algebraically 
solve a particular case: a quartic superluminous dispersion ($\alpha = 2$ in 
eq. \eqref{NL}), in de Sitter space.
We compute the resulting spectrum for all values of $H/\Lambda$.
This analytical treatment allows us to identify the nature of the signatures, and to prepare the numerical
treatment.

In Section~\ref{numerical}, using numerical integration techniques,
we consider the general case, with $\alpha$ ranging from $1$ to $6$.
When $H/\Lambda\ll 1$, we first show that the dominant deviation 
of the power spectrum scales as $(H/\Lambda)^\alpha$.
In other words the leading modification is
\emph{linear} in the lowest order deviation of the dispersion relation.
This is true for both super- and subluminous dispersion,
and for all values of $\alpha$
but for a discrete set of powers ($\alpha_i =3 + 2 i$) 
where an overall $\alpha$-dependent factor vanishes and where the
leading modification scales with a power higher than $\alpha_i$.
Secondly, we show that higher order terms in the 
dispersion relation are irrelevant in that they induce subdominant deviations
which vanish faster than $(H/\Lambda)^\alpha$
in the limit $H/\Lambda \to 0$. 
The signatures are thus robust against 
modifying the dispersion relation by adding higher order terms.
Thirdly, we show that the 
signatures of super- and subluminous dispersion
have equal magnitude and opposite sign
for all $\alpha\neq\alpha_i$, and all values of $H/\Lambda \ll  1$.
Fourthly, when the dispersion relation is smooth and the
asymptotic vacuum well-defined, 
the modifications of the spectrum do not display oscillations.
On the contrary, (fast) oscillations appear when some
non-adiabaticity is localized near a certain (UV) scale,
in agreement with the conclusions of Ref.~\cite{Campo:2007ar}.
We also briefly study the other regime where  $H/\Lambda$ 
is comparable or greater than 1. (For subluminous dispersion
we chose the asymptotic behavior of $F$ so that the 
the adiabatic vacuum is well defined.)
As one might have expected, 
the deviations are no longer governed by the lowest order 
modification of the dispersion relation.
This does not mean however
that this regime should be discarded because it is not known 
at which scale the semi-classical description loses its validity. 
(In some higher dimensional models (see~\cite{Libanov:2005nv} and references therein), 
this scale could be
much smaller than the (4 dimensional) Planck mass, and therefore could be
smaller than $H$.)

We then establish in Section~\ref{slow} that
the modifications of the spectra in slow-roll inflation
can be obtained from the above results. 
The main result is that the 
modifications depend on the wave vector $k$
only through the ratio $H_k/\Lambda$, 
$H_k$ being the value of the Hubble scale when the $k$-mode 
exits the Hubble radius.

In the Conclusions we briefly discuss the possibility of 
distinguishing between modifications of the spectra coming from dispersion and 
other modifications, like those stemming from a change in the inflaton potential. 
In particular, we point that our results predict a violation of the consistency relation, even in the case when 
the scalar and tensor modes obey the same dispersion relation.

\section{\label{modifspec}Modified Power Spectrum}

\subsection{The model}

In this Section, we consider a minimally coupled massless scalar field $\Phi$ 
propagating in a FLRW background spacetime in comoving coordinates:
\be
ds^2 = a(\eta)^2\,(-d\eta^2+d \vec \xi^2)\, .
\ee
The spectrum of scalar perturbations and gravitational
waves can be deduced from the power spectrum of this field, see~\cite{Starobinsky:1979ty,Mukhanov:1992me} and Section~\ref{slow}.
We then introduce a non-linear dispersion relation that we parameterize
by $F$, see eq. \eqref{NL}.
We assume that one recovers the 
standard relation in the IR, i.e. $F^2(P^2) \to P^2 $. We call $\Lambda$ the UV scale
which weighs the first non-linear term of $F^2$.

Decomposing the field in Fourier modes with fixed comoving momentum $k = a P$,
the equation of the rescaled mode $\phi_k = a\Phi_k$ reads
\be
 \left(\partial^2_\eta +
 a^2F^2\left(\frac{k^2}{a^2}\right)-\frac{\partial^2_\eta a}{a}
\right)\phi_k = 0\, .
\label{wave}
\ee
Before considering slow-roll inflation, we first specialize to de Sitter space, 
where 
\be
\left(\partial^2_\eta + \frac{1}{H^2\eta^2}F^2\left(k^2H^2\eta^2\right)-\frac{2}{\eta^2}\right)
\phi_k = 0 \, ,
\ee
since $a= -1/(H \eta)$.
All dependencies in $k$ drop out
when using the variable $x=-k\eta = P/H$:
\be
\left(\partial^2_x + \frac{1}{H^2x^2}F^2(H^2x^2) - \frac{2}{x^2}\right)\phi
 = 0 \,.
\label{wave_x}
\ee
Thus, the power spectrum remains scale invariant in de Sitter space
when the dispersion relation is expressed in terms of proper frequency
and momentum, {\it i.e.}, when it respects the Equivalence Principle,
and when the preferred frame coincides with the cosmological 
frame.~\footnote{There is {\it a priori} no reason for 
the local frame, which governs the violation of Lorentz Invariance in the UV, 
to coincide with the cosmological frame which is associated with the
homogeneous inflationary patch. However, when taking into account the fact that 
the orientation of the local frame should be dynamically
determined~\cite{Jacobson:2000xp},
we expect on general grounds that it will progressively align with the cosmological orientation
as inflation proceeds~\cite{Lim:2004js,Li:2007vz}. However the coupling between the inflaton and a dynamical unit timelike vector field specifying the preferred frame could add a source term to eq. \eqref{wave} and modify non trivially the predictions of inflation, see~\cite{Shankaranarayanan:2005cs}.}

\subsection{Relativistic case}

Let us briefly recall the calculation of the power spectrum for the 
standard relativistic case $F^2 = P^2$.
The mode equation (\ref{wave_x}) reduces to:
\be
\left(\partial^2_x + 1 - \frac{2}{x^2}\right)\phi = 0 \, . \label{wave_x_rel}
\ee
To obtain the power spectrum one needs to identify the asymptotic
positive frequency solution of the above equation.
Indeed, in any successful model of inflation, 
the relevant modes we today observe had  proper
momenta obeying $P \gg H$ at the onset of 
inflation, and  hence were all in their ground 
state~\cite{Parentani:2004ta}.
The power spectrum $P$ of the relevant fluctuations
is thus given by the following~VEV
\ba
P(\eta,k) &=& \frac{k^3}{2\pi^2}\int d^3\xi \, e^{i \vec k\cdot \vec \xi}\, 
\langle 0| \hat\Phi(\eta,\vec \xi)\, \hat\Phi(\eta,\vec 0) |0\rangle
\nonumber \\
 &=& \frac{k^3}{2\pi^2}\vert\,   \Phi^{\rm in}_k(\eta)  \vert^2 \, ,
\label{P_def}
\ea
where $|0\rangle $ is the Bunch-Davies vacuum~\cite{Birrell+Davies} and where $\Phi^{\rm in}_k$
is the positive unit norm Fourier mode associated with this asymptotic state.

The observationally relevant quantity is the value of $P(k)$ at late times, long after horizon exit, 
${k}/{aH}= x \to 0$. When $\Phi^{\rm in}_k(\eta)$
is written in terms of 
\be
u^{\rm in}(x) = \frac{1}{\sqrt{2}}\,\left(1+\frac{i}{x}\right)\,e^{ix}\, ,
\ee
the unit wronskian solution of eq. (\ref{wave_x_rel})
that is purely positive frequency for
$k \eta = - x\to -\infty$,
one obtains
\be
2\pi^2P_0 = H^2 \left(x^2\left|u^{\rm in} \right|^2\right)_{x\to0} 
= \frac{H^2}{2} \, .
\label{P_x_def}
\ee
The index $0$ stands for the unperturbed relativistic case.

Had we worked in 
slow-roll inflation rather than de Sitter space, $P_0(k)$
would have been given by the above with $H$
replaced by $H_k$, the value of $H$ when the $k$-mode exited the
Hubble radius, {\it i.e.}, $k/(a_kH_k) = 1$,
see Section~\ref{slow} for details.

\subsection{\label{cosmo_p_exact}Quartic dispersion relation}

In this subsection, we compute the power spectrum in the particular case
\be
F_{2+}^2 = P^2\left( 1 + \frac{P^2}{\Lambda^2}\right) \, .
\label{Corley}
\ee
The subscript $2$ indicates that the first (and only) non-linearity in $F$ is quadratic in $P/\Lambda$ and the $+$ sign indicates that the dispersion is superluminous.
This dispersion relation~\cite{Corley:1996ar} has been studied 
in a cosmological context in Ref.~\cite{Martin:2002kt}.
However, to our knowledge, the following 
calculation has never been made. 
We believe this is the only dispersive case 
where an exact calculation of the power spectrum
is possible in terms of hypergeometric functions.

\subsubsection{Analytical expression of the power spectrum}
\label{Corley_analytic}
When $F$ is given by eq. (\ref{Corley}), the wave equation becomes:
\be
\left(\partial^2_x + 1 +\frac{x^2}{4\lambda^2} - \frac{2}{x^2}\right)\phi_{2+}
= 0 \, . \label{wave_x_corley}
\ee
From the comparison of this equation with eq. \eqref{wave_x_rel},
one can deduce that the modifications of observables
with respect to the relativistic case 
will all be 
governed by the dimensionless parameter
\be
\lambda = \frac{\Lambda}{2H}\, .
\ee
(The factor $1/2$ has been introduced
for convenience and will be retained throughout the paper.)

As in the relativistic case, the state of the field is chosen to be the 
Bunch-Davies vacuum, that is, 
the ground state associated with the asymptotic positive frequency solution 
of eq. (\ref{wave_x_corley}).
This solution, normalized to unit wronskian, is given by
\be
u^{\rm in}_{2+}(x) = \frac{\sqrt{\lambda} \,e^{-\frac{\pi\lambda}{4}}}{\sqrt{x}} \,
W_{i\frac{\lambda}{2},\frac{3}{4}}\left(-i\frac{x^2}{2\lambda}\right)
\, .
\label{inmode_sup}
\ee
The definition of the Whittaker function $W_{\kappa,\mu}$ can be found 
in~\cite{abramowitz+stegun}, p.505.
That the function $u^{\rm in}_{2+}$ has unit wronskian and is positive frequency 
at early times
can be seen from its asymptotic form for large $x$
(see equation 13.5.2 in~\cite{abramowitz+stegun}):
\be
u^{\rm in}_{2+}(x) \sim 
\sqrt{\frac{\lambda}{x}}\, 
e^{ i\left(\frac{x^2}{4\lambda}+\frac{\lambda}{2}\ln\frac{x^2}{2\lambda}\right)}
 \, .
\label{inmode_sup_early}
\ee
This asymptotic form is precisely the WKB solution of eq. \eqref{wave_x_corley}
with positive frequency.
One verifies that the corrections vanish as $1/x^2$
when $x\to\infty$. Therefore the Bunch-Davies vacuum
is well defined.

The power spectrum is then straightforwardly obtained 
from the behavior of $W_{i\frac{\lambda}{2},\frac{3}{4}}$
for $x \to 0$ (equation 13.5.6 in~\cite{abramowitz+stegun}). 
This gives
\ba
u^{\rm in}_{2+}(x) &=&  
\frac{e^{-\left(\frac{\pi\lambda}{4}
- i \frac{\pi}{8}\right)}}{x}
\,  \sqrt{\frac{\pi}{8}} \, 
\frac{(2\lambda)^{3/4}}
{\Gamma\left(\frac{5}{4}
- i\frac{\lambda}{2}\right)}
\nonumber\\
&&\quad \times\left(1 + O(x^2)\right)
\, .
\label{inmode_sup_late}
\ea
Using this expression we obtain
\ba
2\pi^2 P_{2+}(\lambda) &\equiv & 
H^2\left(x^2 \, \left|u^{\rm in}_{2+}(x)\right|^2\right)_{x\to 0}
\nonumber \\
& & \nonumber\\
&= & 
\frac{H^2}{2}\times\frac{\pi\, (2\lambda)^{3/2}\, 
e^{-\frac{\pi\lambda}{2}}}
{4\, \left|\Gamma\left(\frac{5}{4} -
i\frac{\lambda}{2}\right)\right|^2}\, .
\label{P_sup_exact}
\ea
This expression is exact and valid for all values of $H/\Lambda$.
It gives the power spectrum when using the dispersion relation of eq. \eqref{Corley} 
and when working in the Bunch-Davies vacuum.

Using equation 6.1.45 of~\cite{abramowitz+stegun}, one verifies 
that the factor of the relativistic spectrum $P_0={H^2}/{2}$ 
tends to 1 when $\lambda = {\Lambda}/{2H}\to\infty$. 
In this we recover that the spectrum is \emph{robust}, 
that is, the standard value obtains when adiabaticity is preserved, 
as it is here the case when $H/\Lambda \ll 1$.

\subsubsection{Signatures for small $H/\Lambda$}

Since we have the exact expression of the power spectrum, 
we can properly extract the first corrections in $H/\Lambda = 1/2\lambda$.
When $\lambda\gg 1$, using equation 6.1.47 of~\cite{abramowitz+stegun},
the r.h.s. of \eqref{P_sup_exact} 
can be expanded into:
\be
P_{2+}(\lambda) = P_0
\left(1-\frac{5}{16}\lambda^{-2}+O(\lambda^{-4})\right)
\left(1+e^{-\pi\lambda}\right)\, .
\label{P_sup_approx}
\ee
This expression contains three factors:
the relativistic spectrum and two factors which tend to one when $\lambda \to \infty$.
The first one contains a polynomial starting 
with a quadradic correction in $1/\lambda$,
whereas the correction term of the second is exponentially suppressed. 
In Appendix \ref{wkbtreatment}, we show that 
an approximate calculation based on WKB waves correctly reproduces 
these features, namely that there exist two sources of corrections:
polynomial corrections related to  
local modifications of the mode with respect to the relativistic case,
and exponentially small corrections related to
non-adiabatic transitions.

\subsubsection{Figures}

\begin{figure}
\includegraphics{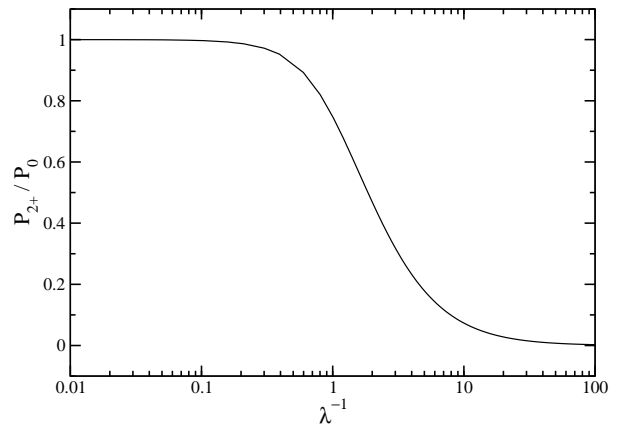}
\caption{\label{fig::CorleyLarge} Plot of the power spectrum, eq. \eqref{P_sup_exact},
for the quartic superluminous dispersion relation eq. \eqref{Corley}
divided by $P_0$, as a function of $\lambda^{-1}= 2H/\Lambda$. 
On the left side, for $\lambda\to \infty$ ($\lambda^{-1}\to 0$), 
the relativistic spectrum is recovered, while for $\lambda\to 0$, 
the power is suppressed as $\lambda^{3/2}$.
}
\end{figure}

To visualize the signatures and to prepare the numerical analysis
of the next Section, we have represented several figures.
In Figure \ref{fig::CorleyLarge} we show $P_{2+}$ of eq. (\ref{P_sup_exact}) divided by $P_0$ 
as a function of 
$\lambda^{-1}$ for $10^{-2} < \lambda^{-1} < 10^{2}$. 
For large $\lambda$, in the adiabatic regime, the modified spectrum
asymptotes to the standard result. Instead,
for small $\lambda$, the modified power spectrum tends to zero as 
\be
P_{2+}(\lambda)\sim  P_0\times  \frac{4\pi}{\Gamma^2(\frac{1}{4})} \,(2\lambda)^{3/2}\, .
\label{P_sup_approx_small}
\ee

\begin{figure}[!b]
\includegraphics{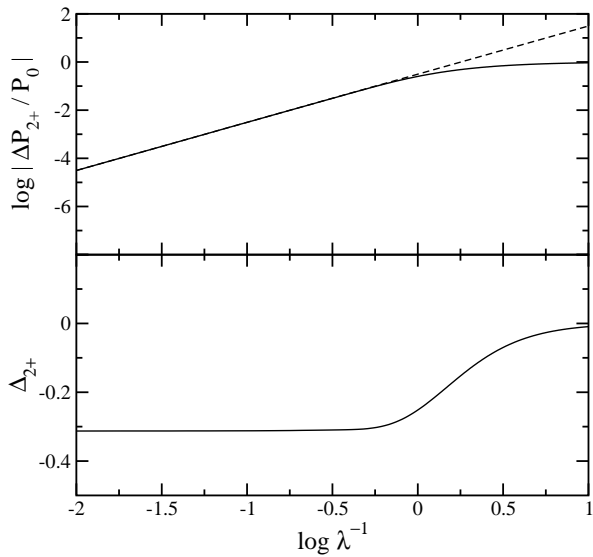}
\caption{\label{fig::CorleyZoom}Plot of the relative corrections 
${\Delta P_{2+}}/{P_0}$, obtained from the exact expression \eqref{P_sup_exact}. In the log-log representation (upper plot), the dominance of the first signature is seen through the 
linearity of the curve: when $\lambda^{-1} < 0.1$
the curve is indistinguishable from the straight line with slope 2 and intercept $\log (5/16) \simeq -0.505$ (dashed line) associated with the quadratic correction in eq. \eqref{P_sup_approx}.
In the lower plot, it is seen through the constantness of $\Delta_{2+}=\lambda^2\Delta P_{2+}/P_0$ for $\lambda^{-1} < 0.1$, with a value equal to the coefficient of the first signature term in \eqref{P_sup_approx}, $-5/16 = -0.3125$.}
\end{figure}

In Figure \ref{fig::CorleyZoom}, to display the signatures of the dispersion, 
we analyze the relative deviation with respect to the relativistic spectrum, 
${\Delta P_{2+}}/{P_0} = ({P_{2+}-P_0})/{P_0}$. 

In the upper plot we have represented 
$\log\left|{\Delta P_{2+}}/{P_0}\right|$ 
as a function of $\log\lambda^{-1}$. 
One clearly sees that when $1/\lambda = 2H/\Lambda < 0.1$,
the polynomial correction term in \eqref{P_sup_approx}
largely dominates all other contributions.
Indeed the curve asymptotes to a straight line, 
with the expected slope, equal to 2, and the expected 
intercept $=\log{5}/{16} \simeq -0.505$. 
In the lower plot, in anticipation to the
numerical analysis, we have represented the quantity
\be
\Delta_{F} (\lambda)
= \lambda^\alpha \, \frac{P_F - P_0}{P_0} \, ,
\label{Delta_def}
\ee
where $\alpha$ is the power of $P/\Lambda$ in the first non-linearity in $F$, $\alpha=2$ here.
The interest of this representation is that as soon as the
deviation in $\lambda^{-\alpha}$ 
becomes the dominant one, the curve becomes horizontal.

\section{\label{numerical}Numerical analysis}

\subsection{\label{disp_rel}Parameterization of the dispersion relation}

In the previous section,
when studying the dispersion relation
with a quadratic 
correction term in $P/\Lambda$, we found that the first deviation of the power spectrum was quadratic in $H/\Lambda$. 
Our first aim is thus to verify whether the power of $H/\Lambda$
characterizing the first deviation is always the same as that of $P/\Lambda$ 
in the leading non-linear
term of the dispersion relation.
Secondly, we want to show that when adding subleading correction terms 
to the
dispersion relation, {\it i.e.}, terms characterized by a  power
of $P/\Lambda$ higher than $\alpha$, 
the signatures are \emph{robust}, {\it i.e.}, they are 
insensitive to these additional terms. Third, we want 
to analyse in a symmetrical manner super- and subluminous dispersion.
Therefore, we select 
subluminous dispersion relations which possess a well defined 
asymptotic regime $P \to \infty$ 
so that the adiabatic vacuum condition can be implemented 
for early times $x=-k\eta \to \infty$. (Remember that subluminous dispersion relations
containing only
one non-linear term
do not possess such a well defined regime, because
$\Omega^2$ becomes negative for $P/\Lambda > 1$.)

To these ends, we consider the following parameterization of the dispersion relation:
\ba
\Om^2 &=& F^2(P,\Lambda; \alpha, \beta, N) 
\label{fullDR}\\
&=& \left\{\beta P + \gamma \Lambda \tanh^{\frac{2}{\alpha}}\left[\left(\frac{\zeta P}{\Lambda}\right)^{\frac{\alpha}{2}}\right]\, 
e^{-\left(\frac{P}{N\Lambda}\right)^{2\alpha+2}}\right\}^2 \, ,
\nonumber
\ea
where the extra parameters $\zeta, \gamma$
verify the following equations:
\ba
\zeta = \left(\frac{3}{4\alpha|1-\beta|}\right)^{1/\alpha}, \quad
\gamma = \frac{1-\beta}{\zeta}\label{gammaConstraint}\, .
\ea
When these are verified, the Taylor expansion of $F^2$ for ${P}/{\Lambda}\ll 1$ is
\ba
F^2 &=& P^2\bigg[1 + {\rm sign}(\beta -1 )\, 
\left(\frac{P}{\Lambda}\right)^{\alpha}\nonumber\\
& & \quad \quad + \tilde N \left(
\frac{P}{\Lambda}\right)^{2\alpha} 
+O\left(\left(\frac{P}{\Lambda}\right)^{2\alpha+2}\right)\bigg]\, ,
\label{OmExpansion}\\
\tilde N &\equiv& 
\frac{1}{4}
\left(1 + \frac{1}{1-\beta}\left(\frac{7\alpha}{10}
+ 1 \right)
\right)\, .\nonumber
\ea
We obtain a superluminous relation when $\beta>1$ and a 
subluminous one when $\beta<1$.
In addition, in each sector,
at fixed $\alpha$, varying $\beta$ modifies only the coefficient of the 
subdominant deviations (with powers of $P/\Lambda$
equal to or greater than $2 \alpha$).

Returning to the exact expression \eqref{fullDR}, we notice
that $\partial_P \ln F \to 0$ 
in the limit $P/\Lambda \to \infty$. Therefore, in all cases 
we reach an adiabatic regime which allows to properly define the asymptotic vacuum, see below
for more details.
The exponential factor in eq. \eqref{fullDR} might seem {\it a priori}
useless since it does not affect any of the first three terms 
in eq. \eqref{OmExpansion}.
It has been introduced to facilitate the numerical integration
because it improves the validity 
of the WKB approximation for large $x$.
The initial vacuum conditions can then be specified for smaller values of $x$
(typically $x_{\rm in} \simeq 500 \lambda/\zeta$), 
thus avoiding a too large accumulation of numerical error
coming from the (physically irrelevant) early evolution
where $P/\Lambda \gg 1$. In the numerical integration
we shall put $N= 100/\zeta = x_{\rm in}/5\lambda $, 
thereby switching on the tanh about two {\it e}-foldings after having 
started the integration. 
(We have checked the 
stability of the results when using higher values of $N$
and $x_{\rm in}/\lambda $.)

\begin{figure*}
\includegraphics{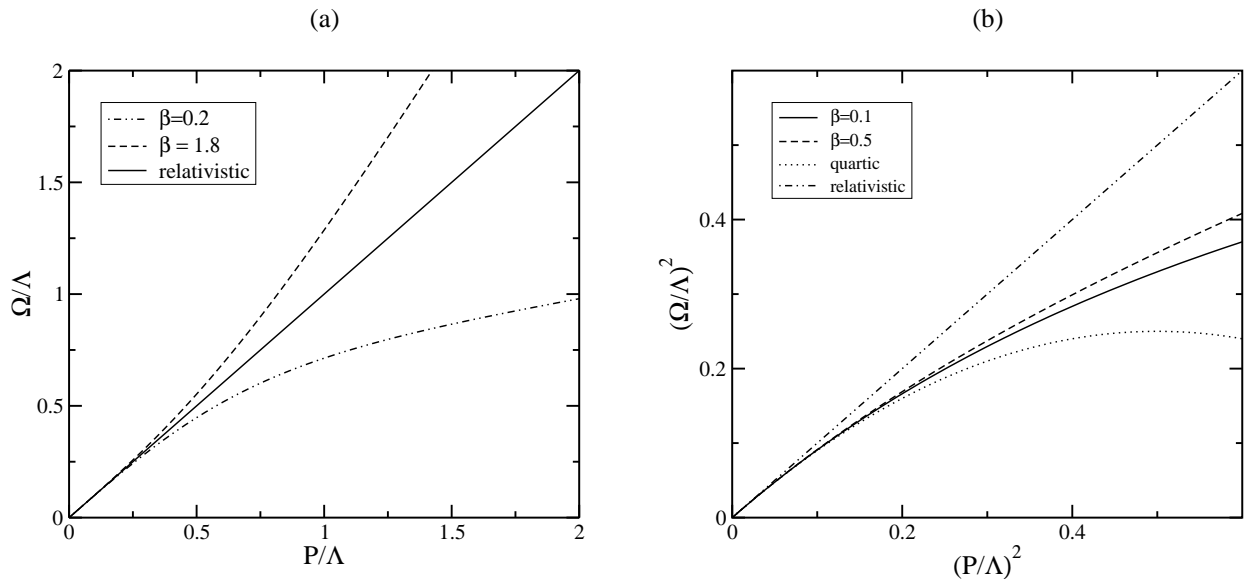}
\caption{\label{fig::fullDR_OmP}Plots of the dispersion relation eq. \eqref{fullDR} for $\alpha=2$ and various values of $\beta$. The linear relativistic dispersion relation is also represented in both plots, as well as the quartic subluminous dispersion relation in the right plot. In plot (b), $\Om^2$ is shown as a function of $P^2$ so that  the quartic dispersion relation be simply a parabola.}
\end{figure*}

Figure \ref{fig::fullDR_OmP} represents the dispersion relation
\eqref{fullDR} for $\alpha=2$ and several choices of $\beta$.
In the left plot the cases $\beta=0.2$ and $\beta=1.8$
illustrate the fact that the parameterization of eq. \eqref{fullDR} encompasses both 
super- and subluminous dispersion relations. In the right plot, the subluminous dispersion relations with $\beta=0.1$ and $\beta=0.5$ are represented along with the quartic dispersion relation 
$F^2 = P^2(1-P^2/\Lambda^2)$. Since the leading (quadratic) 
deviation is the same for all three cases, the curves coincide even after having left the linear regime, until $P\simeq\Lambda/2$.

To study the adiabaticity, we write the corresponding wave equation 
in the variable $x = -k\eta$ as
\be
\left[\partial_x^2 + \om_F
^2\right]\phi(x) = 0\, , \label{wave_fulldr}
\ee
with the effective square frequency
\be
\om_F^2 \equiv \frac{1}{H^2x^2}F^2(H^2x^2,\Lambda;\alpha,\beta, N) 
- \frac{2}{x^2}\, .
\label{om_fulldr}
\ee
The evolution is adiabatic whenever $\sigma_F$ given by
\be
\sigma_F(x)= \vert \frac{\partial_x\ln \om_F}{\om_F}\vert
 \, ,
\label{C}
\ee
remains much smaller than 1~\cite{Brout:1995wp,Niemeyer:2001qe}.

In the absence of dispersion, the non-adiabaticity only 
arises from the $-2/x^2$ term and $\sigma_0$ decreases for $x \gg 1$ as $1/x^3$. 
When adding dispersion, the non-linear behavior of $F$ brings 
another source of non-adiabaticity. However, when $x/\lambda \gg 1$ 
is also satisfied, the non-linearities introduced by 
the hyperbolic tangent in \eqref{fullDR} are 
suppressed by the exponential factor. The 
remaining terms in $\om_F$ 
are a constant term $(=\beta^2)$ and the (small) negative term $-2/x^2$. 
Thus in this asymptotic regime, $\sigma_F$ decreases as $(\beta^2 x^3)^{-1}$ thereby guarantying that the 
adiabatic vacuum stays as well-defined as in the absence of dispersion. 
(Had we not introduced the 
exponential factor, the hyperbolic tangent would asymptote to a constant which would bring a contribution to
$\sigma_F$ decreasing only as $1/x^2$, thereby imposing
to use a larger value of $x_{\rm in}$ to describe with a high precision
the asymptotic vacuum.)

\begin{figure*}
\includegraphics{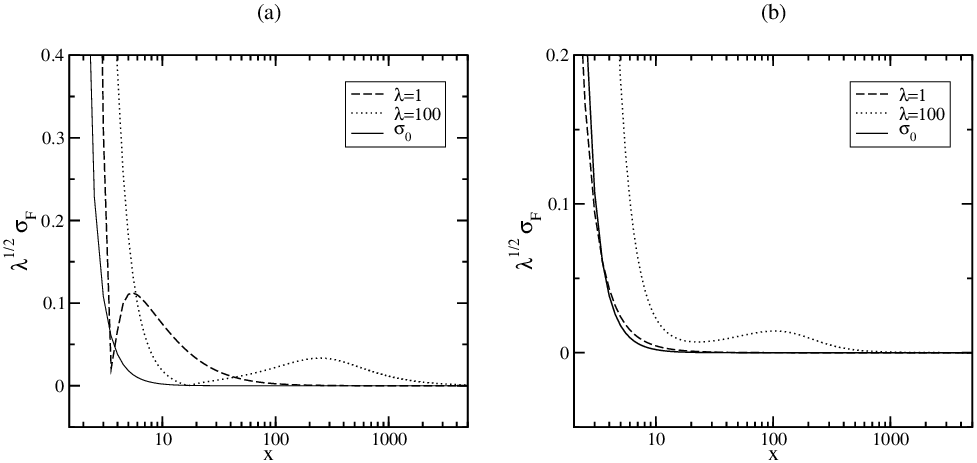}
\caption{\label{fig::fullDR_Cx}Influence of $\lambda$ on the adiabaticity of the evolution.\\$\sigma_F$ is given 
in eq. \eqref{C}.
In all plots $\alpha=2$. To get comparable values we have represented $\sqrt \lambda\,  \sigma_F$. $\sigma_0$ for the relativistic dispersion relation is also shown in both plots for a comparison with the case $\lambda=1$.
(a) Subluminous dispersion ($\beta=0.2$). (b) Superluminous dispersion ($\beta=1.8$). Subluminous dispersion is less adiabatic than superluminous,
as might be expected since $\omega_F$ is reduced in the first 
case. The local maximum at $x\simeq \lambda$ is higher for smaller values of $\lambda$.
}
\end{figure*}

The behavior of $\sigma_F$
is represented in
Figure \ref{fig::fullDR_Cx}
for several values of the parameter $\lambda$ and for subluminous (left plot) and superluminous (right plot) dispersion. The evolution is globally more adiabatic in the superluminous case, as expected since in this case $\om_F$ is greater than in the subluminous case. For each value of $\lambda$, there is a local maximum around $x\simeq \lambda$, whose height diminishes when $\lambda$ increases. When scale separation is not realized ($\lambda = 1$),
in the superluminous case $\alpha=2$ and $\beta=1.8$, this local maximum merges with the rapid increase of $\sigma_F$ when $x$ approaches 1.
In the subluminous case $\alpha=2$ and $\beta=0.2$, $\sigma_F$ takes non-negligible values, of order $10^{-1}$
around $x\simeq 6$, significantly before horizon exit. 

\subsection{Numerical resolution}

Following conventional techniques, the second order differential equation 
\eqref{wave_fulldr} is separated into a system of 4 first order equations, 2 
for the real part of $\phi$ and  its derivative, and 2 for the 
imaginary part of $\phi$ and its derivative. After setting the 
initial conditions, discussed
below before 
eq. \eqref{IC_dphi},
this system is integrated using the embedded 8th order 
Runge-Kutta-Prince-Dormand algorithm provided in the GNU Scientific Library, 
from $x_i$ to some $x_f$ long after the Hubble scale exit ($x=1$),
typically $x_f\simeq 10^{-3}$. 
In this algorithm, the error is estimated at each step and the stepsize adapted to keep the errors within fixed bounds. The use of this algorithm was necessary, because the relative deviation ${\Delta P_F}/{P_0}$ reaches very small values when $\lambda$ is large. For instance it is of order $10^{-8}$ for $\alpha=4$ and $\lambda=100$. 
Since the high values of $\lambda$ also require a larger number $n$ 
of integration points (because $x_i\simeq 500\lambda/\zeta$,
 see below,
is further away), 
the absolute error for each step must be kept under $10^{-10}/n$ if 
we want to be able to distinguish values of $\Delta P_F/P_0$ 
as low as $10^{-10}$.

The initial conditions are fixed using the fact that the WKB solution of 
eq. \eqref{wave_fulldr}, see eq. \eqref{WKBw}, becomes exact when $x \to \infty$. 
For some finite $x_i\gg \lambda$, 
the difference between the WKB solution and the exact positive
frequency solution is of the order of
$\sigma_F$~\cite{Niemeyer:2002kh}.
Thus the WKB approximation becomes excellent whenever 
both 
conditions  $x/\lambda \gg 1$ and $x\gg 1$ are satisfied, 
since in this case
$\sigma_F \sim (\beta^2 x_i^3)^{-1}$.
In practice, the initial conditions are fixed at 
$x_i = 500\lambda$ for $\lambda > 10$ and at $x_i = 5000$ for $\lambda < 10$, so that 
$\sigma_F < 10^{-10}/\beta^2$. 
We thus safely impose
\ba
\phi_F^{\rm in}(x_i)& =& \frac{1}{\sqrt{2\,\om_F(x_i)}}\ ,
\nonumber \\
\partial_x\phi_F^{\rm in}\vert_{x_i} &=& \frac{ i\, \om_F }{\sqrt{2\,\om_F}} 
\left( 1 +  i\,\frac{1}{2}\, \frac{\partial_x \om_F}{\om_F^2}\right)\bigg|_{x_i} \, .
\label{IC_dphi}
\ea
(The arbitrary phase of the 
mode has been put to 0 at $x_i$ 
without affecting the results.)

At the end of the integration of the wave equation, the relative 
deviation of the power spectrum wrt the relativistic spectrum 
is evaluated through:
\be
\frac{\Delta P_F(x_f)}{P_0} 
= 
\frac{P_F(x_f)- P_0}{P_0} 
= 2x_f^2\, |\phi_F^{\rm in}(x_f)|^2 - 1 \, ,
\label{dpop}
\ee
which follows from eq. \eqref{P_x_def}.
Since $x_f$ is not exactly zero, the value obtained contains a 
finite contribution from the decaying mode.
For
$\lambda < 1$, 
this contribution is negligible wrt the modifications due to  
dispersion and we simply ignore it. However, since for $\lambda \gg 1$ it
decreases as $x^2$ and since $x_f\sim 10^{-3}$, the decaying mode contribution in \eqref{dpop} is of order $10^{-6}$. 
It largely dominates the small corrections we are after. 
One could consider smaller values of  $x_f$ 
but this leads to an increase of the number of integration points and of the numerical error.
We thus use the fact that for
large $\lambda$
and $x\ll 1$, $\Delta P_F(x)$ is of the form:
\be
{\Delta P_F(x)} = P_0(A_F + B_F \, x^2)\, ,
\label{P_late}
\ee
where $A_F$ and $B_F$ depend on the values of the various parameters. 
$A_F$ is the contribution of the growing mode that we seek to extract.
To do so, we compute the relative deviation \eqref{dpop} 
at $10x_f$ and $x_f$.
Using twice \eqref{P_late}, 
we properly extract $A_F$.

\subsection{Results}

\subsubsection{Properties of the signatures when $H/\Lambda \ll 1$}

The expansion at large $\lambda$ of the analytical result 
of Section \ref{Corley_analytic},
and the qualitative arguments in Appendix \ref{wkbtreatment}, led us to conjecture that 
$\Delta P_F/P_0$ is linearly related to 
the first deviation in the dispersion relation \eqref{OmExpansion}
and thus possesses the following asymptotic form:
\be
\frac{\Delta P_F}{P_0}\sim \frac{\delta^\pm(\alpha)}{\lambda^{\alpha}}\, ,
\quad 
\textrm{for $\lambda \to \infty$}\, .
\label{dpop_predic}
\ee
Since only the sign of $\beta-1$ appears in the first deviation,
$\delta^{\pm}$ is expected to depend on the superluminous ($+$ exponent) or subluminous ($-$ exponent) character of the dispersion, but not on the 
actual value of $\beta$.

\begin{figure}
\includegraphics{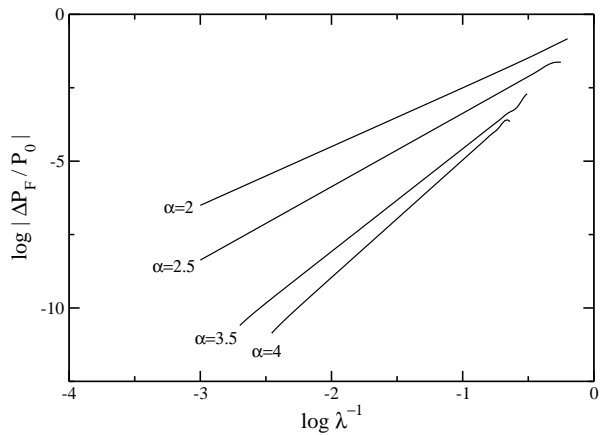}
\caption{\label{fig::betafix_alpha_log}Logarithm of the signature of dispersion for various powers $\alpha$ of $P/\Lambda$. $\beta$ is fixed to 0.2. In each case, the slope is equal to the power $\alpha$, see eq. (\ref{OmExpansion}).}
\end{figure}

In Figure \ref{fig::betafix_alpha_log}, $\log \Delta P_F/P_0$ is represented as a function of $\log \lambda^{-1}$ for a series of subluminous dispersion relations with $\beta = 0.2$ and values of $\alpha$ from 2 to 4.
For high values of $\lambda$ up to $10^3$,
we obtain a set of straight lines, with a slope precisely equal to $\alpha$,
thereby confirming the validity of the asymptotic behavior given 
in eq. (\ref{dpop_predic}).
Had we continued the curves toward $\lambda = 1$, 
we would have seen deviations from this linear behavior. 
This (non-adiabatic) 
regime around and beyond scale crossing will be studied in
Section \ref{scalecrossing}.
For superluminous dispersion we obtain similar results which confirm the 
validity of eq. (\ref{dpop_predic}).

\subsubsection{Robustness of the signatures for $H/\Lambda \ll 1$}

\begin{figure}
\includegraphics{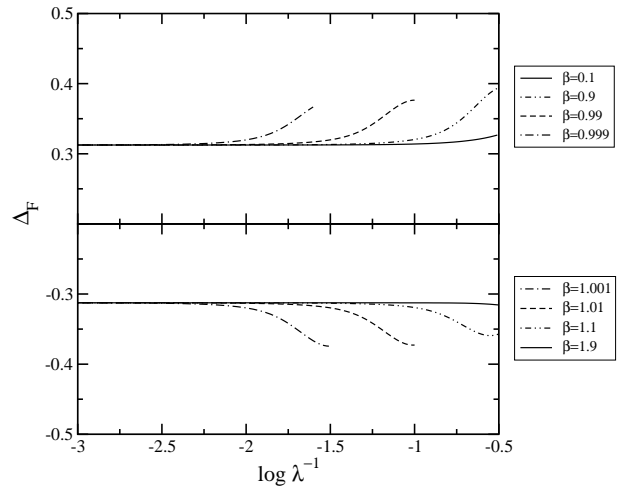}
\caption{\label{fig::alphafix_beta}Robustness of the signatures to higher order terms. $\alpha$ is fixed to 2. One verifies that, both 
for sub- (upper graph) and superluminous dispersion (lower graph), 
the deviations 
of the spectrum induced by 
the second non-linear term (which is weighted
by $(\beta - 1)^{-1}$) disappear for sufficiently large $\lambda$ to give rise 
to the leading deviation given by eq. \eqref{dpop_predic}.}
\end{figure}

To establish that higher order terms in $P/\Lambda$ in the 
dispersion relation are irrelevant in the limit $H/\Lambda\ll 
1$, in Figure \ref{fig::alphafix_beta}
we have plotted $\Delta_F$ of eq. (\ref{Delta_def})
for $\alpha=2$ and various values of 
$\beta$ which weigh the term scaling as $(P/\Lambda)^{2\alpha}$ in 
eq. (\ref{OmExpansion}). The value of $\beta$ ranges
from 0.1, in which case the coefficient of this term 
is of order 1, to $\beta=1 - 10^{-3}$
in which case the coefficient is  
$\simeq 10^3$. Similarly for superluminous dispersion relations, 
$\beta$ ranges from $1 + 10^{-3}$ to $1.9$.
As expected, when $\lambda \gg 1$, all curves agree
irrespectively of the value of $\beta$.
We have numerically verified that the next order deviations ({\it i.e.}
the departure from the asymptotic behavior of eq. \eqref{dpop_predic}
which are clearly seen in Figure~\ref{fig::alphafix_beta})
scale as the square of $(H/\Lambda)^\alpha$. We have also
verified that their normalisation
coefficient contains a contribution which 
is linear 
in $(1-\beta)^{-1}$,
and which comes from the second correction term in eq. \eqref{OmExpansion}, as well as a contribution 
which must arise from the square
of the first correction term in that equation.

Since the function $\Delta_F$ is asymptotically constant when
$\Lambda/H \to \infty$,
the asymptotic properties of the signatures of UV-dispersion 
are governed by the two functions $\delta^{\pm}(\alpha)$ of eq. (\ref{dpop_predic}).
We have verified that similar results hold for all $\alpha$ ranging from 1 to 6, except for $\alpha=3$ and $\alpha=5$, which are particular cases as explained below.

\subsubsection{\label{delta}Properties of $\delta(\alpha)$}

\begin{figure}
\includegraphics{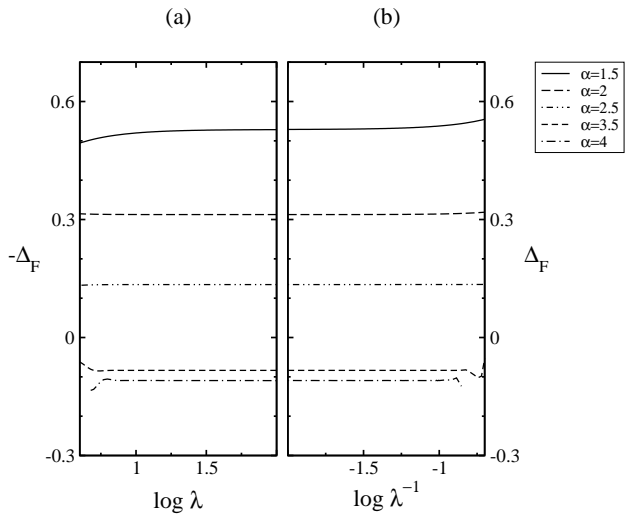}
\caption{\label{fig::betafix_alpha_pow}Comparison of super- and subluminous signatures.\\The legend applies to both plots. In (a) the dispersion is superluminous with $\beta=1.8$. In (b) the dispersion is subluminous with $\beta=0.2$. In (a) we represented $-\Delta_F$ as a function of $\log \lambda$ so that the asymptotic values reached in each plot can be compared at the center of the figure. We clearly see that $\delta^+(\alpha) = -\delta^-(\alpha)$.}
\end{figure}

To further investigate the asymptotic behavior of the signatures,
we have represented in Figure \ref{fig::betafix_alpha_pow}
the function $\Delta_F(\lambda)$ of eq. (\ref{Delta_def})
for various values of $\alpha$ and for
both sub- ($\beta=0.2$) and superluminous 
($\beta=1.8$) dispersion relations.

Besides confirming the fact that $\Delta_F$ is indeed 
constant when $\lambda \to \infty$, we learn from Figure \ref{fig::betafix_alpha_pow} two rather unexpected results.
First, when comparing super- and subluminous cases 
for each value of $\alpha$,
we see that $\delta^+(\alpha)=-\delta^-(\alpha)$.
More precisely, we have numerically checked that 
the sum $\Delta_{\alpha,+}(\lambda)+\Delta_{\alpha,-}(\lambda)$ scales as $\lambda^{-\alpha}$ at large $\lambda$, and thus vanishes when $\lambda\to \infty$.
This result allows us to consider the unique function
\be
\delta(\alpha) = {\rm sign}(1-\beta)\, \delta^\pm(\alpha)
\ee
which does not depend on $\beta$. 
We also point out that we
have not been able to find any analytical
explanation for this simple fact.
See Appendix \ref{wkbtreatment} for an attempt in this sense.

\begin{figure}
\includegraphics{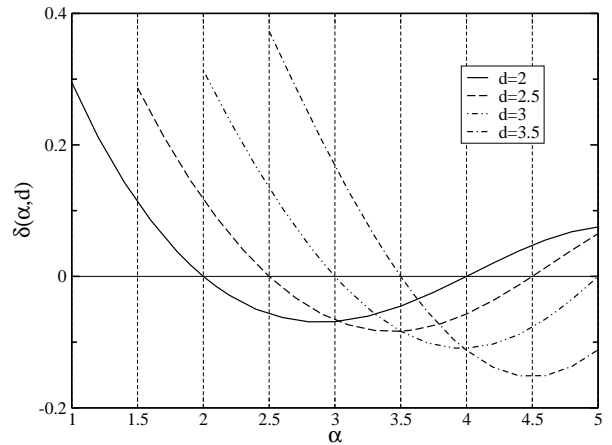}
\caption{\label{fig::delta_alpha}Normalisation $\delta(\alpha,d)$ of the first signature as a function of $\alpha$, for several values of the spatial dimension $d$. For each $d$, $\delta(\alpha,d)$ vanishes and changes sign at the points $\alpha_i = d+2i,\; i=0,1$.}
\end{figure}

Second, we see that the sign of $\delta$ changes between $2.5$ and $3.5$. In fact
it flips sign exactly for $\alpha=3$ when the dimensionality
of spatial sections is $3$.
To establish this second fact, we consider the generalization of eq. \eqref{om_fulldr}
for a de Sitter spacetime of spatial dimension $d$:
\be
\omega_{F,d}^2 = \frac{1}{H^2x^2}F^2(H^2x^2) - \frac{d^2-1}{4x^2}\, .
\label{wave_x_d}
\ee
To obtain $\delta(\alpha,d)$, we constructed a large number of curves $\Delta_{\alpha,d}(\lambda)$ and for each curve found the asymptotic constant value. We repeated this for different values of $d$ 
(including unphysical non-integer values).
The result is shown in Figure \ref{fig::delta_alpha}.
(We verified that, for every $d$, $\delta^+(\alpha,d)=-\delta^-(\alpha,d)$, 
thereby extending the validity of this peculiar correspondance.)

We find that $\delta(\alpha,d)$ vanishes 
precisely when $\alpha$ is equal to $d$.
We also find that after the first zero, it flips sign again precisely at 
$\alpha=d+2$ (for $d=2$ we even verified that it also vanishes at $\alpha=6=d+4$). 
We therefore conjecture that $\delta(\alpha,d)$ changes sign at the points $\alpha_i=d+2i,\; i\in \mathbb{N}$.
In the particular case of the dispersion relation \eqref{Corley}, where $\alpha=2$, the exact calculation of Section \ref{cosmo_p_exact} can be generalized to $d$ spatial dimensions and one finds for large $\lambda$:
\be
\frac{\Delta P_{2+,d}}{P_{0,d}}\sim -\frac{d}{3}\frac{d^2-4}{16}\times \frac{1}{\lambda^2}\, .
\ee
Thus in this case one verifies analytically that the coefficient $\delta^+(2,d)$ vanishes for $d=2$.

\subsubsection{\label{alpha3}Particular case $\alpha=3$ when $d=3$}

\begin{figure}
\includegraphics{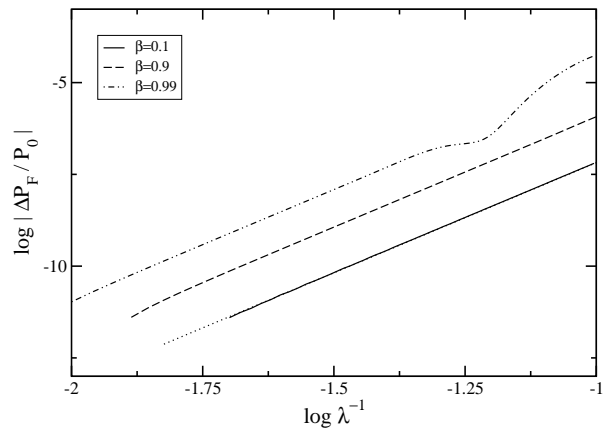}
\caption{\label{fig::alpha3_beta}Logarithm of the relative corrections to the power spectrum for $\alpha=3$ and $d=3$. At large $\Lambda/H$, all curves are linear with the same slope, equal to 6. The dotted line shows the straight line of slope 6 coinciding with the curve $\beta=0.1$ at large $\Lambda/H$.}
\end{figure}

In this subsection we return 
to 3 dimensions and consider the particular case $\alpha=3$ for which $\delta$ vanishes. In Figure \ref{fig::alpha3_beta} we have represented 
$\log \Delta P_F/P_0$ as a function of
$\log\lambda^{-1}$ for $\alpha=3$ and several values of $\beta$.

From the linear  behavior for small $\lambda^{-1}$,
we see that the corrections are still given by a power law.
However, the slope is now $6$, thereby establishing that the
signature is given by the square of $(H/\Lambda)^\alpha$.
This dominant power of $\lambda^{-1}$, {\it i.e.}, the slope, is stable against changes of $\beta$, but its coefficient, related to the intercept, is not.
We verified that this coefficient depends linearly on $(1-\beta)^{-1}$, 
like the coefficient $\tilde N$ of the second correction term in the dispersion relation (see eq. \eqref{OmExpansion}), but that it is not simply proportional to $\tilde N$.
Thus when $\delta$ vanishes,
the dominant deviation of the spectrum
comes both from the square of the first non-linear
term in the dispersion relation and from a linear contribution of the second non-linear term.

\subsubsection{\label{osc}Fast oscillations}

In various works~\cite{Easther:2001fz,Danielsson:2002kx,Niemeyer:2002kh}, 
it was found that the deviations from the standard spectrum display oscillations
with a high frequency proportional to $\Lambda/H$. In these works
the UV scale $\Lambda$ was introduced through the choice of the
initial vacuum which was imposed when $P_{\rm in} = k/a_{\rm in} = \Lambda$
({\it i.e.} $x_{\rm in} = 2 \lambda$ in our notations).
Depending on the adiabaticity of the choice of the vacuum, the 
power in $H/\Lambda$ of the norm of the deviations runs from $1$~\cite{Danielsson:2002kx} to $3$~\cite{Niemeyer:2002kh},
but in all cases, the deviations oscillate with a high frequency.
In Ref.~\cite{Campo:2007ar}, the origin of these oscillations was 
attributed to the instantaneous characterization of the initial state. It was also
shown that the oscillations
are suppressed when smearing the UV scale
(completely or partially depending on the width of $\Lambda$
being larger or smaller than $H$).

Given this, it is interesting to look for oscillations
when the UV scale $\Lambda$ is introduced through a dispersion relation.
In agreement with the conclusions of  Ref.~\cite{Campo:2007ar},
we find that the deviations display no oscillations
when the dispersion relation is smooth,
and when the initial state 
is sufficiently close to the asymptotic adiabatic vacuum. 
When instead the dispersion relation possesses a kink or a bump
near a given UV scale, fast oscillations appear in the 
deviations.~\footnote{Another way to obtain fast oscillations
related to a non-adiabatic evolution has been considered
in Refs.~\cite{Lemoine:2001ar,Brandenberger:2004kx,Danielsson:2004xw}.
In these works, the non-adiabaticity results from the fact that
the proper frequency $\Omega =F(P)$ becomes smaller than $H$ 
for some high momentum $P$.}
We have verified this in several examples.

From this we get two conclusions.
On the one hand, fast oscillations of the type found in Refs.~\cite{Easther:2001fz,Danielsson:2002kx,Niemeyer:2002kh,Martin:2003sg}
can also be obtained with dispersion even when working 
in the asymptotic adiabatic vacuum. On the other hand however, 
to obtain them
requires some odd feature in the dispersion relation well localized near a UV scale 
which will cause a non-adiabatic evolution near that scale.
Thus fast oscillations (with significant amplitude when compared to the deviations)
cannot be considered as a generic property of the 
deviations of the spectra engendered by modifying 
the physics in the UV sector.

\subsection{\label{scalecrossing}Power spectrum for $H/\Lambda \geq 1$}

In our parameterization of the dispersion relation
the asymptotic vacuum is well defined for all values
of $H/\Lambda$. 
Hence we can investigate the properties of the power spectrum
in the much less often studied regime where $H$ is comparable 
and greater than $\Lambda$. (To our knowledge, this was never done before with dispersion, 
see~\cite{Libanov:2005nv,Adamek:2008mp} for an analysis in the presence of dissipative effects.)
It is also of value to note that the simplest approach to 
the trans-Planckian question where there is no dispersion 
but where the quantum state of the field is fixed at some scale $\Lambda$ (see~\cite{Danielsson:2002kx,Niemeyer:2002kh,Martin:2003kp}),
is not defined in the regime $H/\Lambda > 1$.

\begin{figure}
\includegraphics{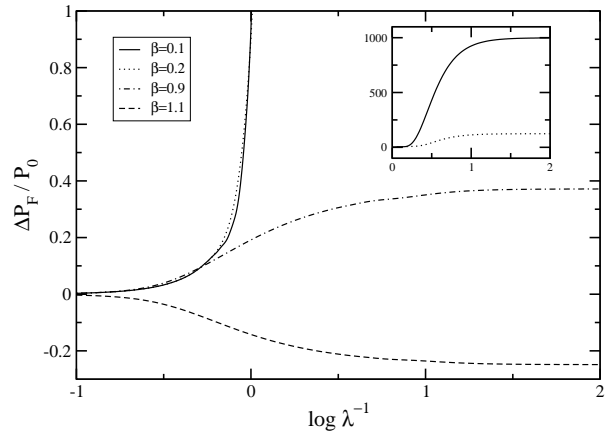}
\caption{\label{fig::scale_crossing}Corrections to the power spectrum 
when scale separation is not realized. 
$\alpha$ is fixed to 2. 
The insert shows that the deviations for $\beta=0.1$ and $\beta=0.2$
are very different for $\lambda^{-1} \gg 1$ even though they agree for $\lambda^{-1} < 1$.}
\end{figure}

Figure \ref{fig::scale_crossing} represents $\Delta P_F/P_0$ as a function of $\log \lambda^{-1}$ for one super- and three
subluminous dispersion relations with $\alpha=2$.
The main plot shows that the different spectra, that 
merge into each other for $\lambda>10$, 
depart from each other and depend strongly on $\lambda$ when $\lambda$ approaches 1.
However, for $\lambda\ll 1$, all curves asymptote to
a constant value that strongly depends on $\beta$.
As already pointed out in Section \ref{disp_rel},
the asymptotic constant behavior
can be understood from 
eq. \eqref{fullDR}:
when $\lambda \ll 1$ there is actually no dispersion before and 
at horizon exit, but the modified speed of the modes is $\beta\neq 1$. 
Thus, when dealing with eq. \eqref{fullDR},
the power spectrum does not depend on $\lambda$ and is equal to $P_0 / \beta^3$  (for all $\beta$, lower or greater than 1), in agreement with what we 
see in Figure \ref{fig::scale_crossing}.

Thus, as expected, when the condition $\lambda \gg 1$ is not satisfied, 
the knowledge of the full dispersion relation is needed to make predictions. 
However, the power spectrum is still well defined as a VEV when the asymptotic vacuum makes sense.
Therefore there is {\it a priori} no reason to exclude the 
possibility that the regime $H \geq \Lambda$ be relevant 
for observational cosmology.
And were this the case, this would complicate 
the identification of the slow roll parameters, not to mention the
reconstruction of the inflaton potential.

\section{\label{slow} Slow-roll inflation}

To show that the above results apply
{\it mutatis mutandis}
to slow-roll inflation, we need to say a few words about the 
kinematics of slow-roll and power law inflation.

A large class of realistic quasi-de Sitter backgrounds can be described in the framework of the slow-roll approximation
where the evolution of the Hubble scale 
is characterized by the smallness of the parameters
\ba
\eps_1 &=& -\frac{1}{H}\partial_t \ln H \ll 1\, ,\\
\eps_2 &=& \frac{1}{2H}\partial_t \ln\eps_1 \ll 1\, .
\ea
The linear order slow-roll approximation consists in keeping only the terms that are linear in $\eps_1$ and $\eps_2$, as well as treating these parameters as constants. This requires that 
the condition
$\partial_t \eps_2 = O(\eps_1^2,\eps_2^2,\eps_1\eps_2)$
be fulfilled. In what follows, expressions in slow-roll inflation 
should be understood as given to linear order in  $\eps_1$ and $\eps_2$.
For the particular case of power-law inflation, $\eps_1$ is constant, $\eps_2=0$,
and the scale factor is exactly given by:
\be
a_{pl} = \left(\frac{1}{-H_0\eta}\right)^{1/(1-\eps_1)}\, .
\label{a_pl}
\ee
When $\eps_2\neq 0$, to linear order 
in the slow-roll parameters, $a_{sl}$ only depends on $\eps_1$:
\be
a_{sl} \simeq \left(\frac{1}{-H_0\eta}\right)^{1+\eps_1}\, ,
\label{a_sl}
\ee
which agrees to (\ref{a_pl})
to first order in $\eps_1$.
This equation follows from
\be
\eta = \int\frac{da}{a^2H} = -\frac{1 + \eps_1}{aH} + O(\eps_1^2,\eps_2^2,\eps_1\eps_2)\, .
\ee 

In slow-roll inflation, 
the time-dependent term in the square frequency of the 
tensor modes is $-\partial^2_\eta a/a$, as in eq. (\ref{wave}),
whereas that of scalar modes is $-\partial_\eta^2f/f$ with $f=a\sqrt \eps_1$~\cite{Mukhanov:1992me}.
In power-law and in first order slow-roll inflation, these
terms are always of the form $-\mu/\eta^2$ where $\mu$ is a constant.
In power-law inflation, the value of the constant is the same for scalar (S) and tensor (T)  modes
and is given by
\be
\mu_{pl} = \frac{2-\eps_1}{(1-\eps_1)^2}\, .
\label{mupl}
\ee
For slow-roll inflation, the values differ and are respectively 
\ba
\mu_{T} &\simeq& 2+3\eps_1\, ,
\label{mutsl} \\
\mu_{S} &\simeq& 2+3(\eps_1+\eps_2) \, .
\label{mussl}
\ea

Given the above discussion, we can deduce the properties of dispersion-induced 
modifications of the spectra 
for scalar or tensor modes in slow-roll inflation,
by simply {\it comparing} the mode equation to
that of  the test field in a de Sitter background. 
Moreover, when considering the leading 
modification of the spectra, since it 
is governed by the lowest order non-linear term of $F$, we can restrict our comparison of the mode equations
to the following. In de Sitter space, in $d$ spatial dimensions, the
relevant terms for a test field are
\be
\left(\partial^2_x + 1 \pm \left(\frac{x}{2\lambda} \right)^{\alpha} 
- \frac{d^2-1}{4 x^2}\right)\phi = 0 \,.
\label{wave_dS}
\ee
In slow-roll inflation, in $3$ dimensions,
using again the variable $x=-k\eta$, they are given by
\be
\left(\partial^2_x + 1 \pm \left(\frac{x}{2\lambda({\eps_1},k)}\right)^{\alpha_{1}} - \frac{\mu}{x^2}\right)\phi_k
 = 0 \,,
\label{wave_sl}
\ee
where the constant $\mu$ is $\mu_{pl}$, $\mu_{T}$ or $\mu_{S}$. 
In power-law inflation, the new parameters are 
\ba
\alpha_{1,pl} &=& \frac{\alpha}{1-\eps_1}\label{alphaeps}\\
2\lambda_{pl}(\eps_1,k) &=& \frac{1}{1-\eps_1}\left(\frac{\Lambda}{H_k}\right)^{1-\eps_1}\label{lambdak}\, ,
\ea
where $H_k$ is 
\be
H_k = \frac{k}{a_{pl}(\eta_k)} 
= \left(\frac{H_0}{1-\eps}\right)^{1/(1-\eps)}\, k^{\eps/(1-\eps)}\, .
\label{Hkeps}
\ee
In first order slow-roll inflation, 
$\lambda_{sl}(\eps_1,k)$ is given by the linearization in $\eps_1$ of eq. 
\eqref{lambdak}.

Comparing equation \eqref{wave_sl} with eq. \eqref{wave_dS}, one sees that the results of the previous sections apply 
with $\lambda$ replaced by $\lambda(\eps_1,k)$, 
$\alpha$ by $\alpha_1$,
and $d$ given by
\be
d_\eps = \sqrt{4\mu+1}\, .\label{deps}
\ee
The condensed notation $d_\eps$ indicates a dependence on both $\eps_1$ and $\eps_2$, and $\mu$ can be $\mu_{pl}$, $\mu_{T}$ or $\mu_{S}$ 
according to the case considered.
Thus, if we denote by $P_{0,\eps}$ and $P_{\alpha,\eps}$ the relativistic and modified spectrum respectively, we 
have, in place of eq. (\ref{dpop_predic}),
\ba
\frac{P_{\alpha,\eps}-P_{0,\eps}}{P_{0,\eps}} &\sim&
\frac{\delta^\pm(\alpha_{1}, d_\eps)}{\lambda(\eps_1,k)^{\alpha_{1}}}\, ,
\label{plmod}
\ea
when $\lambda(\eps_1,k)\to\infty$. 
Using eqs. \eqref{alphaeps} and \eqref{lambdak}, we get
\ba
\frac{P_{\alpha,\eps}-P_{0,\eps}}{P_{0,\eps}} &\sim&
\frac{\delta^\pm(\alpha_{1}, d_\eps)}{(2 -2\eps_1)^{-\alpha_{1}}}
\times \left(\frac{H_k}{\Lambda}\right)^{\alpha}\, ,
\label{plmodbis}
\ea
where the power of $H_k/\Lambda$ {is} $\alpha$, independently of $\eps_1, \eps_2$.

Eq. (\ref{plmodbis}) is the main result of the paper.
For power-law inflation, it
gives the leading modification to the power spectra
to all orders in $\eps_1$.
For slow-roll inflation, it gives the leading modification at linear order in the slow-roll parameters.
It applies both to scalar and tensor perturbations, with different expressions for $d_\eps$, see above.
It establishes that the $k$-dependence of 
the modifications to the spectra only arises through 
$(H_k/\Lambda)^\alpha$.~\footnote{It is interesting to note that 
the modifications we find can be described by the parameterization 
given in eq. (37) of Ref.~\cite{Campo:2007ar}. 
The modifications found in that reference followed from a 
different scheme where there is no dispersion but where
the state of the modes is imposed when $k/(aH_k) = \Lambda$. 
We also notice that the parameter $B$ in eq. (37)
is zero in our case since we find no oscillations.}

Given the above substitutions,
the overall coefficient of $(H_k/\Lambda)^{\alpha}$ 
vanishes for the discrete set of values
\be
\alpha_{i,\eps_1} = d_\eps + 2i\, .
\label{alphais}
\ee
Using \eqref{mupl}, \eqref{mutsl} and \eqref{mussl}, 
one finds
\ba
\alpha_{i,pl} &=& \sqrt{9-6\eps_1+\eps_1^2}+2i\, ,\\
\alpha_{i,T,sl} &=& 3-\eps_1+2i\, ,\\
\alpha_{i,S,sl} &=& 3-\eps_1-\eps_2 + 2i\, .
\ea
For these values, the leading modification scales with a power higher than $\alpha_i$.

A few remarks are in order here. Firstly, as long as 
$H_k/\Lambda \ll 1$, one can extend the correspondance between
de Sitter space and slow-roll inflation to an arbitrary
number of terms of a general dispersion relation.
Indeed, writing the $n^{th}$ term of 
the dispersion relation as
$c_n(P/\Lambda)^{\alpha_n}$, this yields $c_n(x/2\lambda)^{\alpha_n}$ in the wave equation in de Sitter space.
Thus, as above, the corresponding term in slow-roll inflation 
will be given by 
$\lambda\to \lambda(\eps_1,k)$, $\alpha_n\to\alpha_n/(1-\eps_1)$ (or $\alpha_n\to\alpha_n(1+\eps_1)$ in slow roll inflation). 
This correspondance term by term can be used to predict the power of $\lambda(\eps_1,k)$ for an arbitrary 
number of subleading modifications to the spectra. 

Secondly, in the particular cases where the dispersion relation
is of the type \eqref{fullDR} when $N\to\infty$,
i.e. where 
all powers of $P/\Lambda$ are multiples of $\alpha$,
 the \emph{complete} mode equation for 
slow-roll inflation can be obtained by making the above replacements directly in the 
effective frequency of eq. \eqref{om_fulldr}. Thus in this case, the complete modification to the spectra in
slow-roll inflation can be obtained from the de Sitter result, 
for arbitrary values of $H_k/\Lambda$.

\section{Conclusions}

In this paper, we have determined the properties of the modifications of the inflationary power spectra 
engendered by dispersion when the quantum state is the asymptotic vacuum.
When the leading non-linear term in the
dispersion relation is $(P/\Lambda)^\alpha$, see eq. \eqref{OmExpansion},
we have established that the following holds in the regime $H/\Lambda \ll 1$, 
\begin{itemize}
\item The leading modification of the spectrum behaves as 
$\delta(\alpha) \, (H/\Lambda)^\alpha$, for all values of $\alpha$,
except for a discrete set of values where $\delta$ vanishes.

\item 
This leading modification is robust in the sense that when adding
to the dispersion relation
terms containing higher powers of $P/\Lambda$ than $\alpha$, these give rise to subdominant modifications of the power spectrum,
as clearly shown in Figure~\ref{fig::alphafix_beta}.

\item For all values of the power $\alpha$,
except those where $\delta$ vanishes,
the leading modification of the spectrum has equal magnitude and opposite sign when comparing sub- and superluminous dispersion, 
see Figures~\ref{fig::betafix_alpha_pow} and~\ref{fig::delta_alpha}.

\item
When the state is the asymptotic vacuum,
the modifications of the spectra display no oscillations, 
 unless the dispersion relation has some localized bump
that would engender some non-adiabaticity at that scale.

\item These results, which have been derived in de Sitter space,
apply to slow-roll and power-law inflation, both for
scalar and tensor modes, by replacing $H$
by $H_k=k/a_k$ and adjusting some constants 
which appear in the mode equation,
see eq. (\ref{plmodbis}).

\end{itemize}
In brief, when $H_k/\Lambda \ll 1$, when the evolution stays adiabatic, 
and when the state is the asymptotic vacuum, we see that 
it is not necessary to exactly know the 
dispersion relation, since only the first non-linear term matters. 
In particular, in slow-roll inflation, this implies that the dependence on $k$ of the modifications only arises through $(H_k/\Lambda)^\alpha$. 
We also learn that, when $\alpha$ is unknown, 
no general prediction
concerning the properties of the signatures 
of high energy dispersion can be drawn even in the regime $H/\Lambda \ll 1$.

We have also analyzed the `non-adiabatic' regime where $H \geq \Lambda$
that might be relevant in some theories with large extra-dimensions.
The modifications of the spectrum are still well defined
and governed by the dispersion (when working in the
asymptotic vacuum)
even though they are no longer governed
by the lowest order non-linear term of the dispersion relation,
see Figure~\ref{fig::scale_crossing}.

To conclude we discuss to what extent the modifications to the
power spectrum
we obtain could be distinguished, given some observations,
from some other change in the scenario, such as, for instance, a change in
the inflaton potential.

The main point is that the absence of (rapid) oscillations
in eq. \eqref{plmodbis} greatly complicates this identification,
since smooth deformations of the spectra can be accounted for by various means.
As a consequence, the project 
of reconstructing `the' inflaton potential
cannot be pursued without 
making 
some hypothesis concerning the dispersion
relation of the fluctuation modes.

Second, when tensor and scalar fluctuations obey different
dispersion relations, the consistency condition between their spectra is modified, as one might have expected. What is less trivial is that, in slow roll inflation
($\eps_2 \neq 0$),
even if both types of fluctuations are subject to the same dispersion relation, the
modifications of the spectra do not coincide, since the prefactor $\delta$ in eq. \eqref{plmodbis}
depends on the value of the parameter $\mu$,
see eqs.~\eqref{mutsl}, \eqref{mussl}, and \eqref{deps}. 
This remark can be important 
in scenarii where $H\geq \Lambda$.

\begin{acknowledgments}
We would like to thank Martin Bucher, Ted Jacobson,
J\'erôme Martin,  and the referee
for useful remarks.
\end{acknowledgments}

\appendix
\section{\label{wkbtreatment}WKB evaluation of the first corrections}

We show that WKB waves can be used to
reproduce the main features of the analytical solution \eqref{P_sup_exact}, to explain
their physical origin, and to draw predictions
for the corrections  
induced by an
arbitrary dispersion relation. 
However the normalisation of the deviations of the spectrum
cannot be obtained by this method.

The basis of the argument is that the {\it exact} solution of 
eq. \eqref{wave_x} can always be rewritten as a combination of WKB modes:
\be
u(x) = \mathcal{C}(x)\, v(x)+ \mathcal{D}(x) \, \left(v(x)\right)^*\, ,
\label{WKB_exact_decomp}
\ee
where
\ba
v(x) &=& \frac{1}{\sqrt{2\omega(x)}}e^{i\int^x \omega(x')dx'}\, ,\label{WKBw}
\\
\omega(x) &=& \sqrt{\frac{1}{H^2x^2}F^2(H^2x^2) - \frac{2}{x^2}}\, .
\nonumber
\ea
Such a decomposition, however, introduces two unknown functions and is thus underconstrained. 
Following section 2.4 in~\cite{Massar:1997en} we impose:
\be
- i\partial_x u = \om(x)[\mathcal{C}(x) \, v(x)- \mathcal{D}(x)\left(v(x)\right)^*]\, .
\label{Massarcond}
\ee
The unit wronskians of $u$
and $v$ 
together with \eqref{Massarcond} then exactly yields
\be
|\mathcal{C} (x)|^2-|\mathcal{D} (x)|^2=1, 
\label{CmDeq1}
\ee
and 
\be
\partial_x \mathcal{D} = \mathcal{C}(x)\times \frac{\partial_x\om}
{2\om}e^{2i\int^{x}dx'\om(x')}\, .
\label{dxD}
\ee
($\mathcal{C}$ satisfies a similar equation, 
with $\mathcal{C}$ and $\mathcal{D}$ interchanged and a - 
sign in the argument of the exponential.)
Therefore, if in some interval of $x$,
$\sigma=\partial_x \om/\om^2 \to 0$,
$\mathcal{C}(x)$ and $\mathcal{D}(x)$ are approximately constant in it.
We restrict our attention to the dispersion relations such that this is realized for $x\to \infty$.
In this case the {\it in}-mode is simply the asymptotic positive frequency WKB solution,
the asymptotic values of $\mathcal{C}(x)$ and $\mathcal{D}(x)$ 
are $\mathcal{C}=1$ and $\mathcal{D}=0$. We also assume that there is a second interval later on where this condition is realized. Then $\mathcal{D}$ is constant in this interval with a non-zero value $\mathcal{D}_f$. This value represents the probability amplitude of non-adiabatic transitions and is computed to first order by setting $\mathcal{C}=1$ in \eqref{dxD}:
\be
\mathcal{D}_f \simeq -\int_0^{\infty}dx\frac{\partial_x\om}{2\om}e^{
+
2i\int^{x}dx'\om(x') }\, .
\ee
This integral can be evaluated by a saddle point method
when the non-adiabaticity is weak.

We perform this calculation in the particular case of the quartic dispersion relation of equation \eqref{Corley}. 
To start the analysis
we first drop the $-\frac{2}{x^2}$ term in $\om(x)$, which allows us to integrate all the way from $x=+\infty$ to $x=0$. 
The path of integration can be deformed 
continuously into $i\mathbb{R}_+$.
The integral is dominated by the contribution from the saddle point 
$x^*=2i\lambda$ where $\om(x^*)\simeq \sqrt{1+\frac{{x^*}^2}{4\lambda^2}} = 0$. 
Using
\be
\int_0^{x}\sqrt{1+\frac{x'^2}{4\lambda^2}}dx' = \lambda\textrm{arcsinh}\left(\frac{x}{2\lambda}\right)+x\sqrt{1+\frac{x^2}{4\lambda^2}}\, ,
\ee
this yields
\be
|\mathcal{D}_f|= d\,e^{2\textrm{Im}\int_0^{x^*}\om(x)dx} = d\,e^{-\pi\lambda}\, ,
\ee
where the  constant $d$ can be shown to be equal to 1 in the limit 
$\lambda\gg 1$~\cite{Massar:1997en}.
In this limit, $\mathcal{D}_f$ being exponentially 
suppressed, \eqref{CmDeq1} justifies the approximation
 $\mathcal{C}\simeq 1$, and the squared amplitude of the mode for $1\ll |x|\ll \sqrt{\lambda}$ is given by:
\be
|u^{in}(x)|^2\simeq \left|v(x)\right|^2\left(1+2 
\,e^{-\pi\lambda}\cos 2\varphi(x)+ 
\,e^{-2\pi\lambda}\right)\, ,
\label{norm_wkb}
\ee
where $\varphi(x)$ is the phase of $v(x)$ accumulated from
zero to $x$. The important point for us is that $\mathcal{D}_f$ is 
exponentialy suppressed
and not power law suppressed as $\lambda \to \infty$.

We can now qualitatively understand the form of the rhs 
in \eqref{P_sup_approx} by the following argument: neglecting the term $O(e^{-2\pi\lambda})$, before horizon exit, 
but before the WKB approximation completely ceases to be valid, the ratio of the instantaneous values of the power spectrum with and without dispersion  is,
\ba
\frac{P_{2}}{P_0}(x) &\simeq& \frac{|v_2(x)|^2}{|v_{0}(x)|^2}\left(1+2\,e^{-\pi\lambda} \cos2\varphi(x)\right)\nonumber\\
&\simeq& \frac{\om_0(x)}{\om(x)}\left(1+2\,e^{-\pi\lambda}\cos 2\varphi(x)\right)\, ,
\label{corr_Corley}
\ea
where $\om_0(x)^2 = 1 - \frac{2}{x^2}$. 
To first order in $({F^2-P^2})/{P^2}$
we have:
\be
\frac{\om_0(x)}{\om_F(x)} = 1 - \frac{F^2-P^2}{2 P^2}\, .
\label{polycorr_orig}
\ee
For the quartic case, one thus has
\be
\frac{\om_0(x)}{\om(x)} \simeq 1-\frac{x^2}{8 \lambda^2}\, .
\label{polycorr_Corley}
\ee
Thus, eqs. \eqref{polycorr_Corley} and \eqref{corr_Corley} allow us to
explain the origin of the features observed in \eqref{P_sup_approx}:

i) The exponentially small corrections come from the non adiabatic transitions engendered by the dispersion. They are thus the result of a \emph{global} (cumulative) effect. 

ii) The polynomial corrections in \eqref{P_sup_approx} can be viewed
as an imprint of the different normalisation between the relativistic and 
dispersive modes, \emph{around horizon exit}~\footnote{This statement can 
be made precise using the interesting techniques presented in~\cite{Martin:2002vn}: 
one can factorize a singular part in the mode solution so that the WKB approximation
 be applicable to the remaining part after horizon exit. The power spectrum 
 can then be computed in this approximation. 
However, we checked that the 
normalization of the leading deviation 
cannot be obtained in this way. In particular, this approximate calculation cannot explain why this coefficient vanishes for $\alpha=d$.}. These corrections are thus \emph{local} in the sense that they involve only the late time behavior of the modes. The fact that these corrections scale 
as $\lambda^{-2}$ directly  comes
from the power of $\lambda$ in the non relativistic term in the modified dispersion relation.
(In addition, since the normalisation affects equally the positive and 
negative frequency WKB modes, this explains why the polynomial corrections 
are in factor of both the adiabatic and non-adiabatic terms in 
\eqref{P_sup_approx}.) 

We see in \eqref{polycorr_Corley} that the normalisation of the 
superluminous dispersive mode is always smaller than the relativistic normalisation, since $\om_2(x) > \om_0(x)$ for all $x$. 
From this it is tempting to deduce 
that the deviation of the power spectrum is always negative
for superluminous dispersion relations, and 
positive for subluminous ones.
However this conjecture is not correct:
the numerical analysis 
shows that the sign of the deviations flips at $\alpha =3$. 
Therefore the above treatment of the corrections to the WKB approximation
is only indicative when there is mode amplification, that is,
strong departure from adiabaticity.
Thus it cannot be used
to explain why the leading signatures of sub- and superluminous dispersion
found in Figures~\ref{fig::betafix_alpha_pow} have exactly the opposite sign.

However this treatment does contain correct elements. 
To show this,  we consider in the next appendix
a toy model in which the 
UV evolution is identical to that considered here but 
in which there is no horizon exit, 
and therefore no
mode amplification. From the {\it exact} solutions, 
we shall see that only the global exponentially suppressed 
corrections remain, thereby 
demonstrating that the polynomial corrections arise 
near horizon exit and are linked with the mode amplification
(the instability) occuring after horizon exit.

\section{Global corrections and back-scattering}

In this appendix, we consider a toy model where the evolution is governed by the same mode equation as in \eqref{wave_x_corley} 
but with the sign of the non-conformal term $\frac{2}{x^2}$ reversed:
\be
\left(\partial_x^2+1+\frac{x^2}{4\lambda^2}+\frac{2\mu^2}{x^2}\right)\phi = 0 \, .
\label{wave_x_bh}
\ee
we have also slightly generalized the situation by introducing the extra 
parameter $\mu$.

In the present case there is no mode amplification since the
square frequency remains positive. We thus compare 
the pair creation probabilities with and without dispersion, 
and show that only non-adiabatic, exponentially suppressed
corrections appear.

It is worth noticing the relationship between equation \eqref{wave_x_bh} 
and the wave equation for a massless scalar field in a black hole spacetime. 
When studying the radial wave function of s-waves in the 
momentum representation, 
after having factorized out the dependence on 
the Killing frequency (see for instance the appendix of~\cite{Jacobson:2007jx} for details)
 one obtains in the near horizon region the same equation as that
 of eq. (\ref{wave_x_bh}).
Thus we do not exclude that the following exercise be relevant 
for the trans-Planckian signatures in Hawking radiation.

\subsection{Particle creation rate without dispersion}

In the absence of dispersion, the wave equation \eqref{wave_x_bh} reads:
\be
\left(\partial_x^2 + 1+\frac{2\mu^2}{x^2}\right)\phi = 0\, .
\ee
where we restrain ourselves to $\mu>{1}/{\sqrt{8}}$
not to have mode amplification when $x\to 0$.

The asymptotic \emph{in}-mode with positive frequency mode with respect to the time $\eta\propto -x$, 
for $x\to\infty$ is:
\be
\phi^{in} = \frac{\sqrt{\pi}}{2}e^{-\frac{\pi\gamma}{2}}\sqrt{x}H^{(1)}_{i\gamma}(x)\, ,
\label{inmode_rel_noamp}
\ee
where $H^{(1)}_{i\gamma}$ is the Hankel function of the first kind, and $\gamma=\frac{1}{2}\sqrt{8\mu^2-1}$. When $x\to 0^+$, the positive frequency mode is:
\be
\phi^{out} = \sqrt{\frac{\pi}{2\sinh{\pi \gamma}}}\sqrt{x}
J_{i\gamma}(x)\, ,
\label{outmode_rel_noamp}
\ee
where $J_{i\gamma}$ is the Bessel function of the first kind.

We want to compute the Bogoljubov coefficients defined by:
\be
\phi^{in} = \alpha \, \phi^{out}+\beta \, {\phi^{out}}^*\, .
\ee
The squared modulus of $\beta$ gives the particle creation probability.
Using the identity:
\be
H^{(1)}_{i\gamma} = \frac{e^{\frac{\pi\gamma}{2}}}{\sinh\frac{\pi\gamma}{2}}J_{i\gamma}-\frac{1}{\sinh\frac{\pi\gamma}{2}}J_{-i\gamma}\, ,
\ee
we get:
\be
\beta_0 = -\frac{1}{\sqrt{e^{2\pi\gamma}-1}}\, .
\ee

\subsection{Particle creation rate in the presence of dispersion}

We now consider the wave equation \eqref{wave_x_bh}.
The positive frequency {\it in}-mode is:
\be
\phi^{\rm in}_{2+} = \frac{\sqrt\lambda e^{-\frac{\pi\lambda}{4}}}{\sqrt{x}}
W_{i\frac{\lambda}{2},i\frac{\gamma}{2}}
\left(-i\frac{x^2}{2\lambda}\right)
\, ,
\ee
where $\gamma$ is defined as in the previous subsection
and where the Whittaker function
$W$ is the same as in eq. (\ref{inmode_sup}) since both encode
the Bunch-Davies vacuum 
(the value of the second index is different because we have
flipped the sign of the $1/x^2$ term in the mode equation).
The positive frequency {\it out}-mode is:
\be
\phi^{\rm out}_{2+} = \frac{e^{\frac{-\pi\gamma}{4}}}{\sqrt{x}}  \sqrt{\frac{\lambda}{\gamma}}\,
M_{i\frac{\lambda}{2},i\frac{\gamma}{2}}\left(-i\frac{x^2}{2\lambda}\right)
\, .
\ee
where
$M_{i\frac{\lambda}{2},i\frac{\gamma}{2}}$ is another Whittaker function defined in \cite{abramowitz+stegun}.

In this case as well, 
the Bogoljubov coefficients can be directly read from the identity (equation 13.1.34 in \cite{abramowitz+stegun}):
\ba
W_{i\frac{\lambda}{2},i\frac{\gamma}{2}} &=&
\frac{\Gamma(-i\gamma)}{\Gamma(\frac{1}{2}-i\frac{\gamma}{2}-i\frac{\lambda}{2})}
M_{i\frac{\lambda}{2},i\frac{\gamma}{2}}
\nonumber\\
&-& i\frac{\Gamma(i\gamma)}{\Gamma(\frac{1}{2}+i\frac{\gamma}{2} - i\frac{\lambda}{2})} e^{-\frac{\pi\gamma}{2}} \left(M_{i\frac{\lambda}{2},i\frac{\gamma}{2}}\right)^* \, .
\ea
Hence we get
\be
\beta_{2+} = -i\sqrt{\gamma}e^{-\frac{\pi(\lambda+\gamma)}{4}}\frac{\Gamma(i\gamma)}{\Gamma(\frac{1}{2}+i\frac{\gamma}{2}-i\frac{\lambda}{2})} \, .
\ee
The squared modulus simplifies
\ba
|\beta_{2+}|^2 &=& \frac{1}{e^{2\pi\gamma}-1}(1+e^{-\pi\lambda}e^{\pi\gamma})
\nonumber\\
&=& |\beta_0|^2\times (1+e^{-\pi\lambda}e^{\pi\gamma})
\, .
\ea
From the second equation, one immediately sees that in the limit $\lambda\to \infty$ the
pair creation probability in the absence of dispersion is recovered.
And, as announced, in the case when there is no mode amplification, only exponentially suppressed corrections are present, and these correspond to non-adiabatic effects.
A simplified version of this result can be found in a black hole context
in subsection 5.2 of~\cite{Balbinot:2006ua}.

\bibliography{biblio}

\end{document}